\documentstyle[12pt]{article}

\begin{document}
\bibliographystyle{unsrt}

\begin{flushright} UMD-PP-94-50

\today
\end{flushright}

\vspace{6mm}

\begin{center}

{\Large \bf Natural Doublet-Triplet Splitting in Supersymmetric SO(10)
Models\footnote{Work supported in part by a grant from the National Science
Foundation}}\\ [6mm]
\vspace{45mm}

{\bf Dae-Gyu Lee and R. N. Mohapatra}\\
{\it Department of Physics, University of Maryland\\  College Park,
Maryland 20742}\\ [4mm]

\vspace{20mm}

\end {center}

\begin{abstract}
We construct a supersymmetric SO(10) model, where the Dimopoulous-Wilczek
mechanism for doublet-triplet
 splitting is stable under the addition of Higgs superfields
belonging to ${\bf 126 + \overline{126}}$
needed to implement the see-saw mechanism for
neutrino masses and where the charged
 fermion and neutrino mass spectra arise from a
single set of ${\bf {10}}$ and
${\bf \overline{126}}$ Higgs representations.
\end{abstract}

\newpage
In discussing supersymmetric grand unified theories, one has to deal with the
vexing
problem \cite{Ge:82} of doublet-triplet splitting. The problem arises from the
simultaneous
requirement that the $\mbox{SU(2)}_{L}$ doublet and the color triplet
submultiplets of a
Higgs multiplet that generates fermion masses must have very disparate masses:
the
doublet mass must be of order of the electroweak scale $v_{wk}$ to cause
$\mbox{SU(2)}_{L} \times \mbox{U(1)}_{Y}$ breaking whereas the color triplet
mass must
be of order of the GUT scale, $M_{U}$ in order to suppress rapid proton decay.
The
simplest way to achieve this is to fine tune the parameters of the
superpotential. While
such a tree level fine tuning is protected by quantum corrections, thanks to
the
non-renormalization theorem of supersymmetry, it requires unnatural adjustment
of
parameters and is unlikely to be the path chosen by nature. It is somehow more
believable, if underlying group theoretical constraints guarantee the splitting
for arbitrary
choice of parameters of the theory. This possibility is called natural (or
automatic)
doublet-triplet splitting. In this letter, we study this question in the
framework of realistic
SO(10) models. Needless to say that recent indications for neutrino masses in
various
experiments have made SO(10) a more interesting GUT model to consider. It is
therefore
timely to address different aspects of these models.

In the minimal SO(10) models, Higgs superfields belonging to ${\bf 10}$-dim.
representations of SO(10) are used to generate bulk of the quark and lepton
masses. One
important prediction of the dominance of ${\bf 10}$-dim. Higgs in generating
fermion
masses is the equality $m_{b}=m_{\tau}$ at GUT scale. When extrapolated to the
weak
scale, this relation predicts the ratio $m_{b}/m_{\tau}$ in good agreement with
observations. The ${\bf 10}$-dimensional multiplet contains two
$\mbox{SU(2)}_{L}$-doublets (which we denote by $H_{u}$, $H_{d}$) and a color
triplet
and anti-triplet (denoted here as $\xi_{1}$ $(\bf {3})$ + $\xi_{2}$ $(\bf
\overline{3})$). The
problem of doublet-triplet splitting in SO(10) model boils down to
understanding why
$M_{\xi} \simeq M_{U}$ whereas $m_{H_{u,d}} \simeq v_{wk}$. Similar discussions
will
apply to the ${\bf 126}$-dim. representation too, when it plays a role similar
to the $\bf
10$-dim. multiplet.

Several years ago, Dimopoulos and Wilczek \cite{DiWi:82} (DW) suggested a way
to solve
this problem for $\bf 10$-dim. multiplet. They proposed using a ${\bf 45}$-dim.
Higgs
multiplet (denoted by {\bf A}) to break the SO(10)-symmetry by giving vev only
to the
(1,1,15) submultiplet of {\bf A} under $\mbox{SU(2)}_{L} \times
\mbox{SU(2)}_{R} \times
\mbox{SU(4)}_{C}$ subgroup and keeping the (1,3,1) component to have zero vev
naturally. Since {\bf A} is anti-symmetric in the SO(10) indices, the
implementation of this
mechanism requires that there be at least two ${\bf 10}$-Higgs representations
(denoted by
$H_{1}$ and $H_{2}$). It is then clear that, if

\begin{eqnarray}
<A> = \eta \otimes diag (p,p,p,0,0),
\end{eqnarray}
 where
\begin{eqnarray}
\eta \equiv  \left( \begin{array}{cc} 0 & 1 \\ -1 & 0 \end{array} \right),
\nonumber
\end{eqnarray}
a coupling in the superpotential of the form ${\bf A}H_{1}H_{2}$, will make all
the triplets
superheavy while leaving the four doublets
 $H_{1u}$, $H_{1d}$, $H_{2u}$, and $H_{2d}$
light . The subscripts $u$ and $d$
denote the doublets that can couple to up and down type
quarks respectively.
 Since more than two light doublets are known to effect unification of
couplings in an adverse manner,
 one will have to make one pair of these doublets
superheavy. One can solve this problem (a) by letting one of the ${\bf
10}$-Higgs
multiplets (say, $H_{2}$) not couple to fermions and (b) by giving $H_{2}$ a
direct
superheavy mass in the superpotential, i.e., $\mu H_{2} H_{2}$. It has recently
been
argued \cite{BaBa:93} that this may not lead to a strong suppression of proton
decay
though enough to keep the model phenomenologically viable. We will not worry
about this
in order to keep the model simpler.

When one tries to implement the DW idea in realistic models, one immediately
runs into
difficulties. To appreciate this, let us first note that, a SUSY SO(10) model
must have a
${\bf 126 + \overline{126}}$ multiplet pair\cite{AuMo:83}
in order to implement the see-saw mechanism to understand the small neutrino
masses
\cite{GRS:80}. In the presence of the ${\bf 126 + \overline{126}}$ pair
(denoted by $\Delta
+ \overline{\Delta}$), one has the coupling $\Delta A \overline{\Delta}$, which
induces a
vev for the (1,3,1) component of $\bf A$, thereby destroying the
doublet-triplet splitting. If
the coupling is forbidden and the B-L symmetry breaking by $\Delta +
\overline{\Delta}$
vev's is accomplished by a term in the superpotential of the form $(\Delta
\overline{\Delta}
- M^{2})S$ (S being a gauge singlet superfield), then there are a large number
of light
superfields ( due to the SU(126) symmetry of the superpotential), that destroy
the good
unification properties. Something different must therefore be done.

A second requirement that we demand of the minimal SO(10) theory, is inspired
by the
recent observation \cite{BaMo:93} that a single ${\bf 10}$ and single ${\bf
\overline{126}}$
coupling to fermion generations can cure both the problem of bad mass relations
for the
second generation quarks and leptons, i.e., ($m_{s}=m_{\mu}$ and $m_{d}=m_{e}$
at
$M_{U}$) and also generate neutrino masses, provided  the light Higgs doublets
responsible for charged fermion masses is a linear combination of doublets from
the ${\bf
10}$ and ${\bf \overline{126}}$ multiplets. To achieve this mixing, the SO(10)
symmetry
must be broken by a ${\bf 210}$-dim. Higgs multiplet and that there be no extra
light
doublets or color triplet fields left over from the $\Delta$ or
$\overline{\Delta}$ in this
process. It turns out that if the superpotential contains only a term of the
form $\Phi\Delta
H$ (and no $\Phi\overline{\Delta}H$ where $\Phi$ and $H$ respectively denote
the $\bf
210$ and $\bf 10$ dimensional multiplets), this requirement is satisfied.
Combinations of
${\bf 45 + 54}$ often used (e.g. in Ref.~\cite{BaBa:93}) are not adequate for
this purpose.
This is a non-trivial constraint  on model building. Below we present a model
which
satisfies all our requirements, i.e., \\
(a) correct GUT symmetry breaking down to $\mbox{SU(3)}_{C} \times
\mbox{U(1)}_{em}$,\\
(b) only one pair of light Higgs doublets obtained naturally due to the
existence of a
symmetry (perhaps softly broken), \\
(c) the light doublets (both $H_u$ and $H_d$ types) are linear combinations of
the
doublets in the ${\bf 10}$ and ${\bf \overline{126}}$ multiplets.

\vspace*{7mm}
\noindent {\bf The Model}:
We consider the local symmetry group of the model to be SO(10), with an
additional global
symmetry $\mbox{U(1)}_{PQ} \times Z_{16}^{H}$ with both symmetries softly
broken by
some  dimension two terms in the superpotential.

We will demand the symmetry to be respected only by dimension three and higher
terms of
the superpotential and not by the soft dimension two terms.
Thus, there is no light axion in the model. The following set of Higgs
multiplets are chosen
for the model:$\Phi_{1}(210), \Phi_{2}(210), A(45),$ $\Delta(126),$
$\overline{\Delta}(\overline{126}),$ $H_{1,2}(10)$. Their transformation
properties under
the global symmetry groups are given in Table 1. (We denote the sixteenth root
of unity by
$z$.)

The gauge invariant superpotential can be written as a sum of three terms:
\begin{eqnarray}
W=W_{m} + W_{H}^{(3)} + W_{H}^{(2)},
\end{eqnarray}
where
\begin{eqnarray}
W_{H}^{(3)}&=&\lambda_{1} \Phi_{1}^2\Phi_{2} + \lambda_{2} \Phi_{2} \Delta
\overline{\Delta} + \lambda_{3} \Phi_{1} A A + \lambda_{4} A H_{1} H_{2} +
\lambda_{5}
\Phi_{1} \Delta H_{1}, \\
W_{H}^{(2)}&=&\mu_{1} \Phi_{1}^2 + \mu_{2} \Phi_{2}^{2} + \mu_{3} \Delta
\overline{\Delta} + \mu_{4} A A + \mu_{5} H_{2} H_{2} + \mu_{6} \Phi_{1}
\Phi_{2}, \\
W_{m}&=&h_{ab} \Psi_{a} \Psi_{b} H_{1}.
\end{eqnarray}

We choose all main parameters $\mu_i$ to be of order of the GUT scale $M_U$.
The two
terms $W_{H}^{(3)}$ + $W_{H}^{(2)}$ lead to a  number of degenerate
supersymmetric
minima, one of which has the desired pattern of symmetry breaking with $<A>$
given in
Eq. (1) naturally with the vev of $\Phi_{1}$ only along the (1,1,1) and
(1,1,15) directions
and naturally with that of  $\Phi_{2}$ along the (1,1,1), (1,1,15), and
(1,3,15) directions. By
the Dimopoulos-Wilczek mechanism, this leads to natural doublet-triplet
splitting for both
$H_1$ and $H_2$; the $\mu_5$ term makes one pair of the light doublets
superheavy as
required phenomenologically.

To show that this pattern of vev's preserves supersymmetry down to the
electroweak
scale, we call $<\Phi_{i}(1,1,1)>=a_{i}$, $<\Phi_{i}(1,1,15)>=b_{i}$,
$<\Phi_{i}(1,3,15)>=c_{i}$, $<A(1,1,15)>=p$, $<A(1,3,1)>=q$,
$<\Delta(1,3,\overline{10})>=$ $<\overline{\Delta}(1,3,10)>=v_R$. All these
vev's  are of
order $M_U$.
Let us first write down the vanishing F-term conditions for the $<A>$; we get
\begin{eqnarray}
2 \mu_4 p + 2 \lambda_{3} {\sqrt{2} \over 3}  p b_{1} + 2 \lambda_{3} {1
\over{\sqrt{6}}} q
c_{1} =0, \nonumber \\
2 \mu_4 q + 2 \lambda_{3} {1 \over{\sqrt{6}}} q a_{1} + 2 \lambda_{3} {1
\over{\sqrt{6}}} p
c_{1} =0.
\end{eqnarray}
{}From this, we first see that there is a solution of Eq.~(6) for which $p \neq
0$ and $q=0$ if
$c_{1}=0$. This is the vacuum we will focus on and see if the rest of the
F-term conditions
are satisfied for arbitrary values of the parameters in the superpotential. To
see this, let us
write down \cite{HeMe:90} those $F=0$ conditions with $c_{1}=0$ and $q=0$.
\begin{eqnarray}
0 &=& 2 \mu_{1} a_{1} + \mu_{6} a_{2}, \nonumber \\
0 &=& 2 \mu_{1} b_{1} + \mu_{6} b_{2} + 2 \lambda_{1} {b_{1} b_{2} \over{9
\sqrt{2}}} +
\lambda_{3} {\sqrt{2} \over 3} p^2, \nonumber  \\
0 &=& \mu_{6} c_{2} + 2 \lambda_{1} ({a_{1} c_{2} \over{6 \sqrt{6}}} + {b_{1}
c_{2} \over{9
\sqrt{2}}}).   \\
0 &=& 2 \mu_{2} a_{2} + \mu_{6} a_{1} + \lambda_{2} {v_R^2 \over{10 \sqrt{6}}},
\nonumber \\
0 &=& 2 \mu_{2} b_{2} + \mu_{6} b_{1} + \lambda_{1} {b_{1}^2 \over{9 \sqrt{2}}}
+
\lambda_{2} {v_R^2 \over{10 \sqrt{2}}},  \nonumber \\
0 &=& 2 \mu_{2} c_{2} + \lambda_{2} {v_R^2 \over{10}}.  \\
0 &=& 2 \mu_{2} v_R + \lambda_{2} ( {a_{2} v_R \over{10 \sqrt{6}}} + {b_{2} v_R
\over{10
\sqrt{2}}} + {c_{2} v_R \over{10}}).
\end{eqnarray}

Eqs.~(7)-(9) arise from the F-terms corresponding to $\Phi_{1}$, $\Phi_{2}$,
and $\Delta$
(or $\overline{\Delta}$), respectively. It is easy to see that this has
non-trivial solutions for
all the vacuum expectation values for arbitrary choice of the parameters of the
model. This
establishes the naturalness of the doublet-triplet splitting.

Let us now write down the doublet Higgsino mass matrix and isolate the massless
pair of
the Higgs doublets, that generate electroweak symmetry breaking. First we note
that the
doublets in $H_{2}$ pick up superheavy mass due to the $\mu_{5}$ term and are
completely decoupled from the other doublets. Denoting the rest of the doublets
by
$(\Phi_{2u}, \mbox{ } \Phi_{1u}, \mbox{ } \Delta_{u}, \mbox{ }
\overline{\Delta}_{u}, \mbox{ }
H_{1u})$ and $(\Phi_{2d}, \mbox{ } \Phi_{1d}, \mbox{ } \overline{\Delta}_{d},
\mbox{ }
\Delta_d, \mbox{ } H_{1d})$, we can write down their mass matrix in the basis
where the
above sets of fields denote the columns and rows respectively.

\begin{eqnarray}
\left( \begin{array}{ccccc} \mu_{2} &\tilde{\mu}_{6} & 0
 & \lambda_{2} v_{R} & 0 \\    \tilde{\mu}_{6} &
\tilde{\mu}_{1} & 0 & 0 & \lambda_{5} v_{R} \\ 0 & 0 & \tilde{\mu}_{3} & 0 & 0
\\
\lambda_{2} v_{R} & 0 & 0 & \tilde{\mu}_{3} & \lambda_{5} v_{U} \\  0 & 0 &
\lambda_{5}
v_{U} & 0 & 0 \end{array} \right). \end{eqnarray}

Analyzing this matrix, we conclude that it has two zero eigenstates given by:
\begin{eqnarray}
H_{u} &=& x_{1} H_{1u} + x_{2} \overline{\Delta}_{u} + x_{3} \Phi_{1u} + x_{4}
\Phi_{2u},
\nonumber \\
H_{d} &=& y_{1} H_{1d} + y_{2} \overline{\Delta}_{d}.
\end{eqnarray}

We therefore see that the light doublets have the desired property to solve the
charged
fermion mass puzzles of the simplest SO(10) models following
Ref.~\cite{BaMo:93}. All
colored triplets in this model are superheavy - thus, proton decay
is suppressed.

Let us now turn to the fermion sector. Note that $W_{m}$ in Eq. (5) has two
deficiencies:
(a) it does not give correct relation between $m_{s}$ and $m_{\mu}$, (b) the
absence of
the $\overline{\Delta}$ coupling makes the right-handed neutrinos massless.
Both these
problems are cured when we include the $Z_{16}^{H}$-invariant Planck scale
induced
dimension 4 terms in the superpotential, i.e.,
\begin{eqnarray}
W_{m}^{(1)} = {1 \over{M_{Pl}}} f_{ab}\Psi_{a}\Psi_{b}\Phi \overline{\Delta}.
\end{eqnarray}

A {\em priori}, there are three independent couplings, where the $\Psi \Psi$
bilinear
transforms like $\bf 10$, $\bf 120$, $\bf 126$; however, the fermion masses get
contribution only from $\bf 10$, $\bf \overline{126}$ type couplings. The $\bf
10$ type
couplings simply redefine the original $\bf 10$ couplings, whereas the $\bf
\overline{126}$
type couplings introduce the same structure to the fermion mass as in
Ref.~\cite{BaMo:93}
and predictions of neutrino masses given in that reference carry over.
If $f_{ab}$ are chosen to be order 1 to 3 and the $\overline{\Delta}$ -vev's
along the
$(2,2,15)$ direction are chosen to be of order 100 GeV, the contribution of
$W_{m}^{(1)}$
to fermion masses have the correct order of magnitude. Secondly, the
right-handed
neutrinos acquire a mass of order $\simeq 10^{12}$ GeV or so if $v_{R} \simeq
10^{-1}v_{U}$. This scale for the right-handed neutrinos is helpful in
understanding
baryogenesis of the universe \cite{FuYa:86}.

\vspace*{7mm}
In summary, we have constructed an SO(10) model where  the Dimopoulos-Wilczek
mechanism for doublet-triplet splitting is unaffected by the addition of ${\bf
126 +
\overline{126}}$ multiplets needed for the see-saw mechanism for neutrino
masses. We
have also been able to show that once the lowest order Planck scale corrections
are
included, the light Higgs doublets have the correct structure to reproduce the
desired
quark-lepton mass spectra and lead to predictions for neutrino masses as in
Ref.~\cite{BaMo:93}.
There are no undesirable light particles in the theory. We hope to have made
clear the nontrivial nature of the constraints one would like to impose on
a desirable SO(10) model and that by choosing appropriate symmetries, models
satisfying all these constraints can indeed be constructed. Perhaps, the
existence of one such model demonstrated in this letter will stimulate further
thinking to
construct
simpler models that may yield further insight into SO(10) grandunified
theories.  \\
\noindent We thank K. S. Babu for some discussions.

\section*{Table Caption}
Table 1: Global symmetry quantum numbers for different fields. $z=e^{i\pi/8}$.

\vspace*{10mm}
{\small
\begin{center}

\begin{tabular}{ |c|c|c| } \hline\hline
Supermultiplets &$\mbox{U(1)}_{PQ}$ charge &$Z_{16}^{H}$ \\ \hline
Matter multiplets &             & \\
$\Psi_{3}$         &+1  &$z^{7}$  \\

$\Psi_{2}$         &+1  &$z^{7}$  \\

$\Psi_{1}$         &+1  &$z^{7}$  \\ \hline
Higgs multiplets &             & \\
$\Phi_{1}$          &0   &$z^{12}$  \\
$\Phi_{2}$          &0   &$z^{8}$  \\
 A                  &0   &$z^{2}$    \\
$\Delta$            &+2  &$z^{2}$    \\
$\overline{\Delta}$    &$-2$  &$z^{6}$    \\
$H_{1}$             &$-2$     &$z^{2}$    \\
$H_{2}$             &+2     &$z^{12}$    \\ \hline\hline
\end{tabular}
\end{center}
\begin{center}
Table 1

\end{center}
}

\end{document}